\documentclass[utf8]{frontiersFPHY} % for Physics and Applied Mathematics and Statistics articles

\setcitestyle{square} % for Physics and Applied Mathematics and Statistics articles
\usepackage{url,hyperref,lineno,microtype,subcaption}
\usepackage[onehalfspacing]{setspace}

\usepackage{graphicx}% Include figure files
\usepackage{bm}% bold math
\usepackage{amsmath,amssymb}
\usepackage{siunitx} % SI units

\newcommand{\ie}{\emph{i.e.}}
\newcommand{\eg}{\emph{e.g.}}

\newcommand{\kBT}{{k_\te B T}}
\newcommand{\kB}{{k_\te B}}
\newcommand{\HBM}{\text{HBM}}

\newcommand{\te}[1]{\mathrm{#1}}
\newcommand{\df}{\mathrm{d}}
\newcommand{\vv}[1]{\boldsymbol{#1}}
\newcommand{\ev}[1]{\langle #1 \rangle}
\newcommand{\MYabs}[1]{\lvert #1 \rvert}

\def\keyFont{\fontsize{8}{11}\helveticabold }
\def\firstAuthorLast{Sven Auschra {et~al.}} %use et al only if is more than 1 author
\def\Authors{Sven Auschra\,$^{1,*}$, Dipanjan Chakraborty\,$^{2}$, Gianmaria Falasco$^{3}$, Richard Pfaller\,$^{1}$ and Klaus Kroy\,$^{1}$}
\def\Address{$^{1}$Institute for Theoretical Physics, 
University of Leipzig, 
04103 Leipzig, Germany
\\
$^{2}$ Department of Physics, Indian Institute of Science Education and Research, Punjab, India
\\
$^{3}$Department of Physics and Materials Science, 
University of Luxembourg, 
L-1511 Luxembourg, Luxembourg
}
\def\corrAuthor{Klaus Kroy}

\def\corrEmail{klaus.kroy@uni-leipzig.de}

\begin{document}
\onecolumn
\firstpage{1}

\title[Coarse Graining Nonisothermal Microswimmer Suspensions]{Coarse Graining Nonisothermal Microswimmer Suspensions} 

\author[\firstAuthorLast ]{\Authors} %This field will be automatically populated
\address{} %This field will be automatically populated
\correspondance{} %This field will be automatically populated

\extraAuth{}% If there are more than 1 corresponding author, comment this line and uncomment the next one.
%\extraAuth{corresponding Author2 \\ Laboratory X2, Institute X2, Department X2, Organization X2, Street X2, City X2 , State XX2 (only USA, Canada and Australia), Zip Code2, X2 Country X2, email2@uni2.edu}

\maketitle

\begin{abstract}

  We investigate coarse-grained models of suspended self-thermophoretic microswimmers. Upon heating, the Janus spheres, with hemispheres made of different materials, induce a heterogeneous local solvent temperature that causes the self-phoretic particle propulsion.
  Starting from atomistic molecular dynamics simulations, we verify the coarse-grained description of the fluid in terms of a local molecular temperature field, and its role for the particle's thermophoretic self-propulsion and hot Brownian motion.
  The latter is governed by effective nonequilibrium temperatures, which are measured from simulations by confining the particle position and orientation. They are theoretically shown to remain relevant for any further spatial coarse-graining towards a  hydrodynamic description of the entire suspension as a homogeneous complex fluid.

\tiny
\keyFont{%
  \section{Keywords:} homogenisation, active particles, microswimmers, hot Brownian motion, non-isothermal molecular dynamics simulations
}
\end{abstract}

\section{Introduction}

\label{sec:Intro}

Mesoscale phenomena are at the core of current research in hard and soft matter systems~\cite{Aharony2010perspectMesoPhysics,Kroy2015focusSoftMeso}.
The reason for this is at least twofold.
Firstly, some of the most interesting states of matter are not properties of single atoms or elementary particles, but emerge from many-body interactions, at the mesoscale; e.g., the mechanical strength of many materials is determined by low-dimensional mesostructures.
Secondly, these interesting mesoscale properties are often insensitive to molecular details and amenable to widely applicable coarse-grained models that provide both physical insight and efficient control~\cite{Laughlin2000theoryOfEverything}.
Extensive atomistic computer simulations can therefore usually be bypassed either by much more efficient coarse-grained numerical techniques \cite{gompper2009mpc,carenza2019LatticeBoltzmann,shaebani2020ComputModels} or even by analytical methods \cite{steffenoni2016InteractingBD,laemmel2017MesoSandTransport}.
Both exploit the universality of the mesoscale physics to compute experimental observables without having to resolve the atomistic details.
The price one pays for this efficiency is that fluctuations, which are increasingly important in biophysical and nanotechnological applications \cite{Bregulla2014stochLocal,Jikeli2015spermNavi,Qian2013photonNudging,selmke2018nudging1,selmke2018nudging2}, may get renormalized or even inadvertently lost upon coarse graining.
It is then not always obvious how they have to be properly re-introduced when need arises \cite{Adhikari12005fluctLattBoltz}.
Systems with non-equilibrium mesoscale fluctuations, such as suspensions of self-propelled particles and other active fluids \cite{menon2010ActiveMatter,marchetti2013hydrSoftActMat}, are of particular interest in this respect.

One might imagine an approach based on non-equilibrium thermodynamics, which, like hydrodynamics itself, is often valid down to the nanoscale, if judiciously applied \cite{Celani2012anomTD}.
But this theory's starting point is a macroscopic deterministic one, without fluctuations, so that it is natively blind to the refinements we are after.
The framework of stochastic thermodynamics would seem more appropriate, but, in its current formulations, temperature gradients, which are of particular interest to us, are explicitly excluded \cite{Seifert2012stochTD}.
So the question that we address here, namely how nonisothermal and other non-homogeneous fluctuations scale under hydrodynamic coarse-graining, is not only of practical interest, but is also a profound theoretical problem that affects the construction of hydrodynamic theories, in general.

Our strategy is to start from a complete atomistic description of a well-defined model system that allows for analytical progress, yet provides the basis for simulating a number of innovative technologies \cite{Braun2013optThermophorTrap,Qian2013photonNudging,Bregulla2014stochLocal,selmke2018nudging1,selmke2018nudging2}.
The system is a solvent of Lennard--Jones atoms with embedded nanoparticles that are themselves made of Lennard--Jones atoms but maintained in a solid state by additional FENE attractions.
The computer simulation of the model reveals that, even upon mesoscopic heating, nanoparticles and solvent admit a local-equilibrium description in which a (molecular) temperature field $T(\vv r, t)$ can be defined that represents the \emph{local molecular temperature} at position $\vv r$ and time $t$ almost down to the atomic scale.
In other words, the notion of a rapid local thermalization of the molecular degrees of freedom, in the conventional sense of canonical equilibrium, is still reasonable, even for very small volume elements.
As it turns out though, important hydrodynamic degrees of freedom of the system are, in general, not locally thermalized at $T(\vv r, t)$, unless this field is everywhere equal to the constant ambient temperature $T_0$ (in which case the fluid is in a global isothermal equilibrium).
In other words, there is \emph{a priori} no obvious recipe how to canonically construct an ``average molecular temperature'' that would allow us to start from the atomistic model and move up to a coarse-grained description by some straightforward low-pass filtering.
In the following we demonstrate what can be done, instead, and why and how the resulting coarse-grained model deviates from naive expectations.

\begin{figure}[tb!]
  \centering
  \includegraphics[width=0.3\columnwidth]{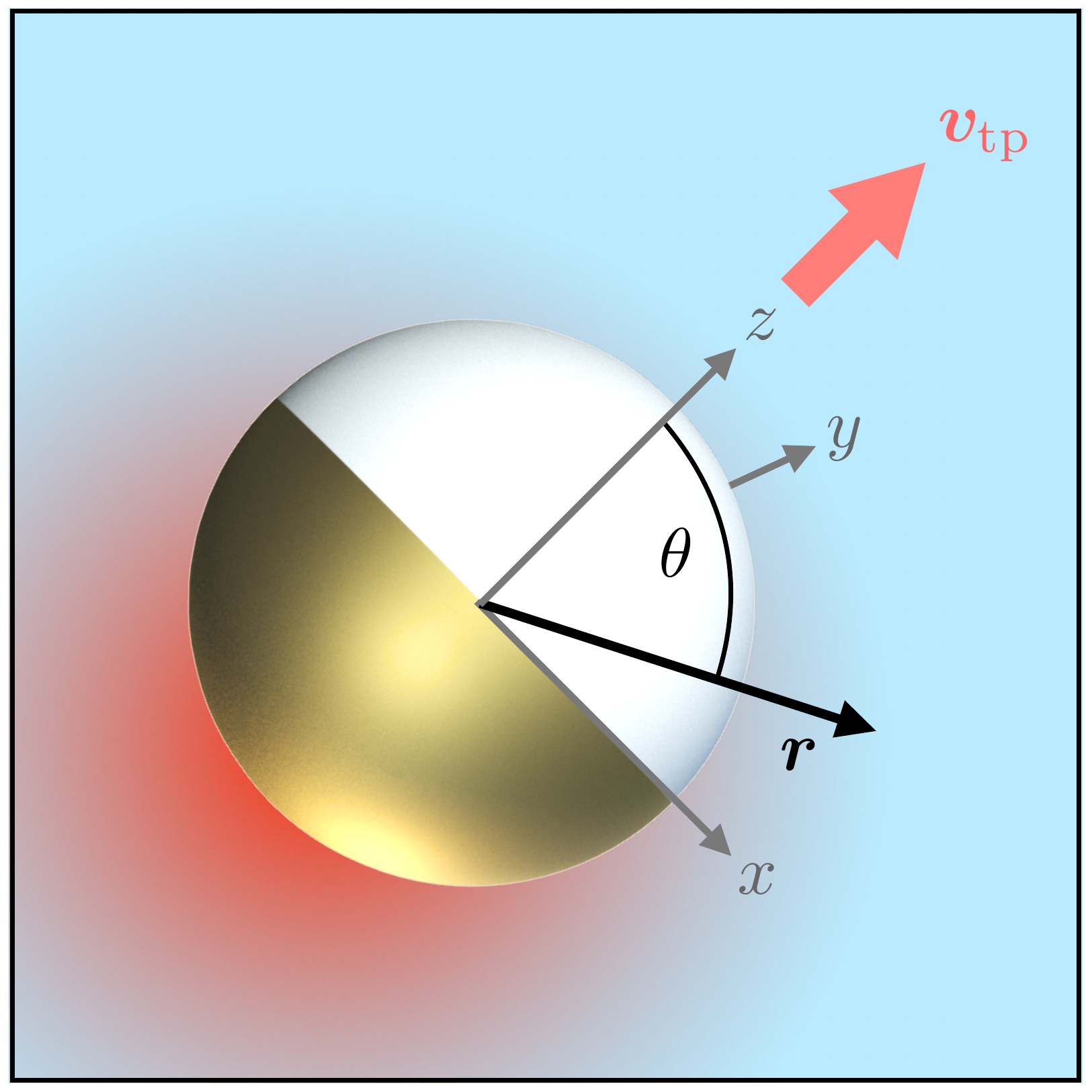}
  \caption{Spherical Janus particle coated with a thin gold layer on one hemisphere. Upon heating, the particle induces an anisotropic temperature profile in the ambient fluid. The resulting thermoosmotic interfacial flux gives rise to a net propulsion at swim speed $v_\te{tp}$ along the symmetry axis \cite{Jiang2010selfTPofJP,bregulla2015PhozoSwimmers,kroy2016HotMicroswimmers}. A reference frame attached to the particle's geometric center has its $z$-axis  aligned with the symmetry axis, pointing towards the uncoated hemisphere. The polar angle between the position vector $\vv r$ and the $z$-axis is denoted by $\theta$.}
  \label{fig:setup}
\end{figure}

The paper is organized as follows. Section~\ref{sec:an-atom-impl-1} introduces the atomistic description of our model. The first coarse-graining step that admits the formulation of a local temperature field $T(\vv r, t)$ is done in Sec.~\ref{sec:molec-temp}. Section~\ref{sec:HBM} reviews some basic results from the theory of hot Brownian motion that permits a first-principle calculation of the Brownian fluctuations of a thermally homogeneous nanoparticle exposed to the field $T(\vv r, t)$, and therefore provides the basis for a theory of nonisothermal Brownian motion.
Section~\ref{sec:THBM_JanusSphere} considers the more challenging case of a thermally anisotropic particle, specifically a Janus particle as depicted in Fig.~\ref{fig:setup} (alongside some notation).
Not only does the symmetry breaking complicate the computation of its hot Brownian fluctuations compared to an isotropic particle, it also gives rise to a spontaneous anistropic solvent flow in its vicinity \cite{Anderson1989collTransp,Bickel2013flowPatternJP}. Such a particle therefore advances thermophoretically along its symmetry axis, under non-isothermal conditions \cite{Jiang2010selfTPofJP,bregulla2015PhozoSwimmers}. In the supplemental material pertaining to this article, we provide evidence that our simulation method indeed leads to a well-controlled, sizable net propulsion of the Janus particle, as was already reported in \cite{kroy2016HotMicroswimmers}.
 Finally, in Sec.~\ref{sec:CoarseGraining} we address the largely open task of homogenizing a whole suspension of hot, active particles, before we close with a brief conclusion.

\section{Atomistic model of a hot Janus swimmer}
\label{sec:an-atom-impl-1}

We consider a heated metal-capped Janus sphere immersed in a fluid as depicted in Fig.~\ref{fig:setup}. 
In order to resolve microscopic details, such as the interfacial thermal resistance and the mechanism of thermophoresis \cite{Anderson1989collTransp}, our simulation is based on a schematic molecular model, in which both the fluid and the Janus particle are atomistically resolved.
All atomic interactions are modeled by a modified Lennard--Jones 12--6 potential
\begin{equation}
  \label{eq:LJ_wetting_potential}
  U_{\alpha \beta}(r) = 4 \epsilon
  \left[
    \left(
      \frac{\sigma}{r}
    \right)^{12}
    - c_{\alpha \beta}
    \left(
      \frac{\sigma}{r}
    \right)^{6}
  \right],
\end{equation}
(truncated at $ r = \SI{2.5}{\ensuremath{\mathnormal{\sigma}}}$). We henceforth measure length, energies and times in terms of the Lennard--Jones units $\sigma$, $\epsilon$, and $\tau \equiv \sqrt{m \sigma^2/\epsilon}$, respectively, where $m$ denotes the atomic mass.
The solvent molecules always interact via the standard Lennard-Jones potential, \ie, $c_{\alpha \beta} = c_{\text{ss}} = 1$. The Janus particle constitutes a spherical cluster of Lennard--Jones atoms additionally bound together by a FENE potential
\(
U(r)
=
-0.5\kappa R_0^2 \ln
  \left[
    1 - \left( r/R_0 \right)^2
  \right],
\)
with the spring constant $\kappa = \SI{30}{\ensuremath{\mathnormal{\epsilon / \sigma^2}}}$ and $R_0 = \SI{1.5}{\ensuremath{\mathnormal{\sigma}}}$ justified in \cite{grest1986mdSimPolymers}.  The prefactor $c_{\alpha \beta}$ in Eq.~\eqref{eq:LJ_wetting_potential} determines the position $(2 \sigma^6 / c_{\alpha \beta})^{1/6}$ of the interaction potential's minimum.  By varying $c_{\alpha\beta}$ we can tune the wetting properties of the nanoparticle surface and thereby its Kapitza heat resistance~\cite{Barrat1999wetting,Barrat2003kapitzaRes}.  
To mimic the anisotropic physio-chemical properties of Janus spheres, typically realized in experiments \cite{Jiang2010selfTPofJP,bregulla2015PhozoSwimmers} by capping one hemisphere of a polystyrene (p) bead with a thin gold (g) layer, we
employ the parameters $c_{\text{gs}}$ and $c_{\text{ps}}$, representing gold-solvent and polystyrene-solvent interactions, respectively. The gold cap is modelled by a $\SI{1}{\ensuremath{\mathnormal{\sigma}}}$-layer of Lennard--Jones particles on one hemisphere.
In order to achieve sizeable thermophoretic motion, the imperfect heat conduction between the swimmer's surface and the adjacent fluid due to the Kapitza heat resistance is important \cite{kroy2016HotMicroswimmers}.  In our simulations, a periodic cell of length $L = \SI{50}{\ensuremath{\mathnormal{\sigma}}}$ was filled with roughly $10^5$ solvent molecules, and the Janus sphere, which itself is composed of 200 atoms constituting a particle of radius $R \approx \SI{3.2}{\ensuremath{\mathnormal{\sigma}}}$.
With an integration time step of $\SI{0.005}{\tau}$, our simulations then proceed as follows: First, the system is equilibrated in the $NPT$ ensemble using a Nose-Hoover thermostat and barostat at a temperature of $T_0 = \SI{0.75}{\ensuremath{\mathnormal{\epsilon / k_{\text{B}}}}}$ and a thermodynamic pressure of $p = \SI{0.01}{\ensuremath{\mathnormal{\sigma^3/\epsilon}}}$ to ensure equilibration of the Lennard-Jones fluid into a liquid state~\cite{Errington2003quantifyOrderLJ,Potoff1998critPointLJ}.
In the subsequent heating phase, we apply a momentum conserving velocity rescaling procedure to thermostat the gold cap atoms at a temperature $T_{\text{P}} > T_0$ while keeping the solvent atoms at the boundaries of the simulation box at $T_0$. The described procedure indeed realizes sizable self-propulsion of the Janus particle \cite{kroy2016HotMicroswimmers}. Its propulsion speed and direction are determined by the wetting parameters $c_{\alpha \beta}$ of the interaction potential \eqref{eq:LJ_wetting_potential} and the heating temperature $T_{\te P}$ of the gold cap as shown in the supplemental material.

Colloidal thermophoresis has been studied extensively  by means of mesoscopic theories and atomistic computer simulations \cite{ganti2017MDThermoSlip,burelbach2018colloidThermophor,burelbach2018thermophorForces,proesmans2019thermoosomis}. The following sections focus on a specific aspect, namely  the enhanced thermal fluctuations experienced by a heated Brownian particle in its (self-created) nonisothermal environment. The swimmer's so-called hot Brownian motion inevitably interferes with its self-propulsion randomizing particle position and orientation. In the following section, the crucial elementary notion for theories of hot Brownian motion, nameley that of a molecular temperature field at which the Lennard--Jones fluid locally equilibrates, is properly introduced, analytically studied, and tested against simulation results.

\section{Molecular Temperature Field}
\label{sec:molec-temp}

\begin{figure}
  \centering 
  \includegraphics[width=0.5\columnwidth]{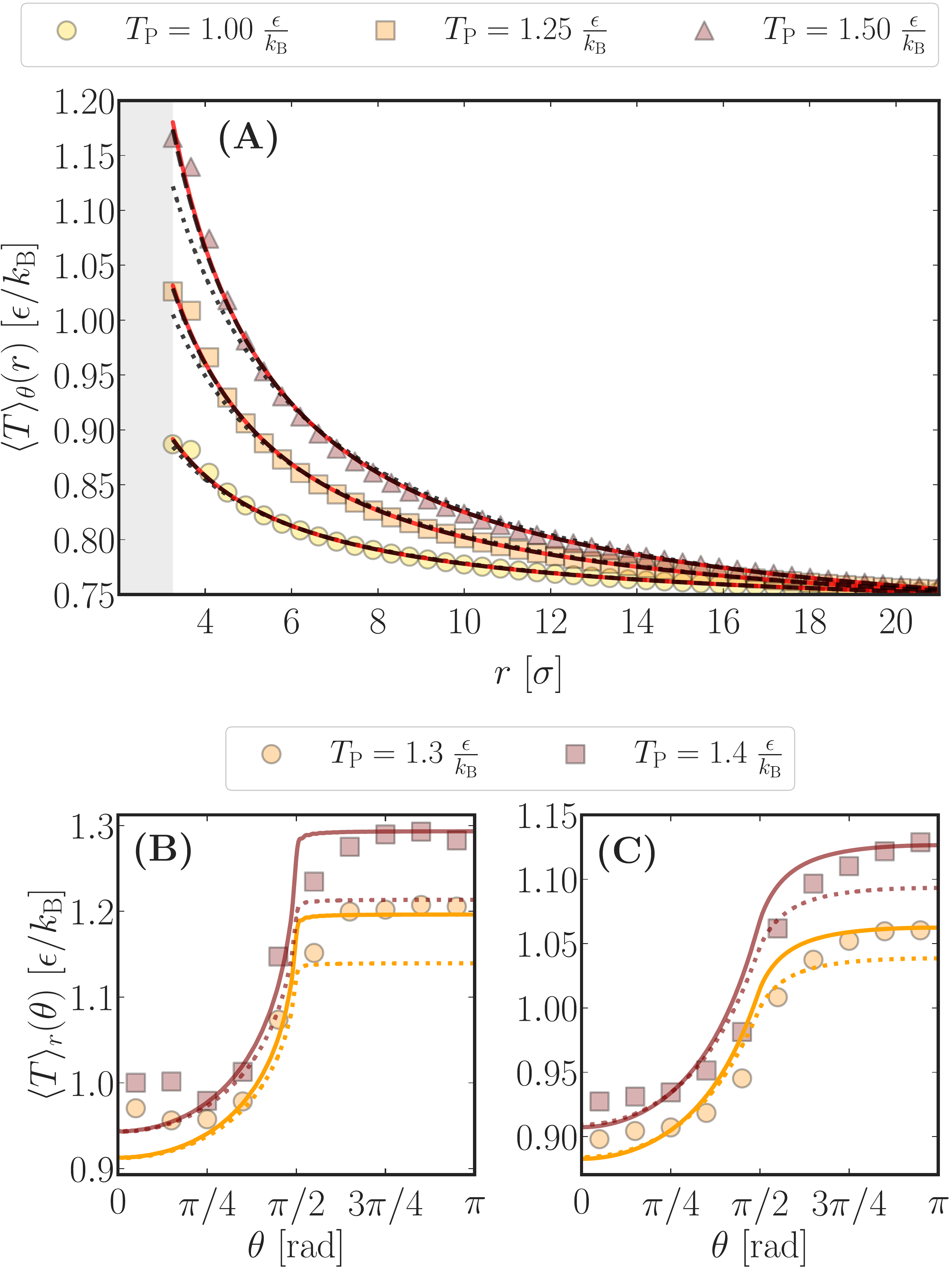}
  \caption{Mean fluid temperatures extracted from MD simulations (symbols) with wetting parameters $c_{\te{gs}}=2$ and $c_{\te{ps}}=1$ compared to theory (curves). \textbf{(A)} Mean fluid temperature at radius $r$ averaged over thin concentric spherical shells of thickness $\SI{0.0844}{\sigma}$ for three different heating temperatures $T_{\te{P}}$ of the gold cap.
    The Janus spheres' surface is indicated by the gray bar.
    Using $T_0$ and $\Delta T$ as fit parameters, theory curves are obtained from Eq.~\eqref{eq:mean_T_of_r} (dotted black), $\theta$-averaging the dipole approximation of Eq.~\eqref{eq:T_thick_cap_kappa=1/T} (solid red), and Eq.~\eqref{eq:T_hom} (dashed black). For the latter, the mean temperature increment $\ev{\Delta T}_\theta$ served as fit parameter.  \textbf{(B)/(C)} The measured radially averaged fluid temperature as a function of the polar angle for two distinct heating temperatures,  averaged up to a distance of $\SI{0.1}{\sigma}$ and $\SI{2.5}{\sigma}$ from the surface, respectively.  For the dotted theory curves we radially averaged the first 100 terms of Eq.~\eqref{eq:Tfield_thick_cap_const_kappa}, while the solid lines were obtained from the first 100 terms of the numerically averaged profile in \eqref{eq:T_thick_cap_kappa=1/T}; parameters $T_0$, $\Delta T$ as obtained in \textbf{(A)}.}
  \label{fig:Tprofile}
\end{figure}

In order to justify the concept of a \emph{molecular temperature field} $T(\vv r, t)$, we measured the average kinetic energy of the fluid atoms within (i) thin concentric spherical shells of radial thickness $\approx\SI{0.1}{\sigma}$ around the particle's geometric center and (ii) angular bins of size $\pi/10$ around the Janus sphere. Due to the symmetric particle composition, we expect the temperature profile to depend only on the radial distance $r$ from the particle's geometric center and the polar angle $\theta$ with respect to its symmetry axis. Figure \ref{fig:setup} supports these considerations visually.
The corresponding angle-averaged and radially averaged temperature profiles, $\ev{T}_\theta(r)$ and $\ev{T}_r(\theta)$, are  presented in Fig.~\ref{fig:Tprofile} for various heating temperatures $T_{\rm P}$ of the gold shell.
In accordance with previous studies regarding isotropic, homogeneously heated Brownian particles \cite{Chakraborty2011generalEinsteinRel,Rings2012rotHBM,Rings2011theoryOfHBM}, we find the temperature profiles to be stationary, as expected from the strong time scale separation between the swimmer's motion and the kinetic and energetic equilibration in the solvent. We exploit this fact in Sec.~\ref{sec:HBM} to estimate the effective temperatures governing the particle's enhanced hot Brownian motion.

We start our discussion with the  thermodynamic description of heat conduction.  In steady state, the heat conduction equation for the temperature profile $T(\vv r)$ reads \cite{LandauLifschitz10}
\begin{align}
  q(\vv r)
  &=
    \vv \nabla
    \cdot
    \left[
    \kappa(\vv r) \vv \nabla T(\vv r)
    \right]
  \\
  \label{eq:heat_equation_general}  
  &=
    \vv \nabla \kappa(\vv r)
    \cdot
    \vv \nabla T(\vv r)
    +
    \kappa(\vv r)
    \vv \nabla^2 T(\vv r),
\end{align}
where $\kappa(\vv r)$ denotes the heat conductivity and $q(\vv r)$ is the heat flux absorbed by the gold cap.
Equation \eqref{eq:heat_equation_general} is accompanied by boundary conditions at the particle surface $r=R$.
In our simulations, a sudden temperature drop  at the particle-fluid interface signifies a substantial Kapitza heat resistance. In the simplified theory, we neglect this effect and enforce the continuity of the temperature profiles inside and outside the particle, 
\(
T_{\te{in}}(R,\theta)
=
T_{\te{out}}(R,\theta)
\),
for the sake of simplicity.   Following the derivations in \cite{Bickel2013flowPatternJP}, we also demand continuity in the normal component of the heat flux,
\begin{equation}
  \label{eq:heat_flux_cont_uncapped}
  \kappa_{\te{in}} \partial_r T_{\te{in}}
  =
  \kappa_{\te{out}} \partial_r T_{\te{out}}
  \quad
  \te{for}
  \quad
  0 \leq \theta \leq \pi/2,
\end{equation}
along the uncoated part of the Janus sphere.  Motivated by its very large heat conductivity, the gold cap of the Janus sphere is modelled as an isotherm kept at surface temperature
\begin{equation}
  \label{eq:thick-cap}
  T(R,\theta) = T_0 + \Delta T
  \quad
  \te{for}
  \quad
  \pi/2 \leq \theta \leq \pi,  
\end{equation}
where $\Delta T$ denotes the increment relative to the ambient temperature $T_0$ of the fluid. We further set the heat conductivities
\(
\kappa_{\te{in}}
=
\kappa_{\te{out}}
\equiv
\kappa
\)
for the remainder of this section. Besides, we henceforth omit the subscript for the outer temperature profile, $T_\te{out} \equiv T$, as the temperature profile inside the particle is not relevant in the following. If, as a first crude approximation, we finally assume that $\kappa = const.$ throughout the whole system, the temperature field $T(r,\theta)$ is given by \cite{Bickel2013flowPatternJP}
\begin{equation}
  \label{eq:Tfield_thick_cap_const_kappa}
  T(r,\theta)
  =
  T_0
  +
  \Delta T
  \sum\limits_{n=0}^\infty
  B_n P_n(\cos\theta)
  \left(
    \frac{R}{r}
  \right)^{n+1},
\end{equation}
with the Legende polynomials $P_n$ and expansion coefficients
\begin{equation}
  \label{eq:Legnde_Coeffs}
  B_0
  =
  \frac12
  +
  \frac{1}{\pi},
  \quad \quad
  B_{2k}
  =
  -B_{2k+1}
  =
  \frac{1}{\pi}
  \cdot
  \frac{(-1)^k}{2k+1}.
\end{equation}
Due to the orthogonality relations ($\delta_{lk}$ denotes the Kronecker-delta)
\begin{equation}
  \label{eq:Pn_orth}
  \int\limits_{-1}^1 \df c ~
  P_k(c) P_l(c)
  =
  \frac{2}{2l +1} \delta_{kl},
\end{equation}
and with the short-hand notation $c \equiv \cos\theta$, the angle-averaged temperature profile $\ev{T}_\theta(r)$ simplifies to 

\begin{align}
  \label{eq:mean_T_of_r_first}
  T_0 + \ev{\Delta T}_\theta
  &=
    \frac{
    \int_0^\pi \df \theta \,
    T(r, \theta) \sin \theta
    }{
    \int_0^\pi \df \theta \, \sin \theta
    }
    = 
    \frac{
    \int_{-1}^1 \df c \, T(r, c)
    }{
    \int_{-1}^1 \df c
    },
  \\
  &=
    T_0
    +
    \frac{\Delta T}{2}
    \sum\limits_{n=0}^\infty
    \frac{R^{n+1}}{r^{n+1}}
    \int_{-1}^1 \df c \,
    B_n P_n(c),
  \\
  \label{eq:mean_T_of_r}
  &=
    T_0
    +
    \Delta T B_0
    \frac{R}{r}.
\end{align}
We infer from Eq.~\eqref{eq:mean_T_of_r} that the actual and  mean surface temperature increment are related via 
\begin{equation}
  \label{eq:B0}
  B_0
  =
  \frac{\ev{\Delta T}_\theta}{\Delta T}.
\end{equation}
Using the ambient fluid temperature $T_0$ and the surface increment $\Delta T$ as parameters, we fitted Eq.~\eqref{eq:mean_T_of_r} to the numerically simulated average temperature profiles.  The resulting fits are presented in Fig.~\ref{fig:Tprofile} (A) for three distinct heating temperatures.  Matching the measured temperature profiles well at larger distances, the theory curves slightly but systematically underestimate the numerical data closer to the particle surface.  As a result, the mean surface temperature increment $\ev{\Delta T}_\theta$ and thus, via Eq.~\eqref{eq:B0}, also $\Delta T$ itself, are underestimated.  This becomes even more pronounced when comparing measurements of the radially averaged temperature data $\ev{T}_r(\theta)$ to the corresponding radial average of the theory curve \eqref{eq:Tfield_thick_cap_const_kappa} as shown in Fig.~\ref{fig:Tprofile} (B) and (C).  The former plot depicts $\ev{T}_r(\theta)$ close to the particle surface, whereas for the latter one, we averaged up to a distance $\SI{2.5}{\sigma}$ away from the swimmer's surface.  In both cases the dotted theory curves lie significantly below the numerical data, especially on the coated part ($ \pi/2 \leq \theta \leq \pi$) of the particle.

The described shortcomings of the theoretical temperature profile \eqref{eq:Tfield_thick_cap_const_kappa} can be improved by taking the temperature dependence of the heat conductivity $\kappa(T)$ of the fluid into account.
For the studied Lennard--Jones fluid, the heat conductivity approximately follows a $\left( T^{-1} \right)$-law \cite{Chakraborty2011generalEinsteinRel}.  Plugging $\kappa \propto 1/T$ into Eq.~\eqref{eq:heat_equation_general}, the ansatz
\(
T(r,\theta)
=
T_0
\exp[\psi(r,\theta)]
\)
yields the equation
\(
\vv \nabla^2 \psi(r,\theta) \propto q(r, \theta)
\)
for the auxiliary dimensionless field $\psi(r,\theta)$.
Sticking to thick-cap boundary conditions, \eqref{eq:heat_flux_cont_uncapped} and \eqref{eq:thick-cap}, $\psi(r,\theta)$ solves
the same boundary value problem as the temperature field in the previously discussed case of $\kappa = const.$, albeit the latter boundary condition now reads
\(
\psi(R,\theta)
=
\ln
\left(
  1
  +
  \Delta T/T_0
\right)
\)
for $\pi/2 \leq \theta \leq \pi$. 
% \begin{align}
%   \label{Laplace_psi}
%   \vv \nabla^2 \psi(r,\theta)
%   &=
%     0,
%   \\[0.5em]
%   \label{eq:psi_continuity}
%   \psi_{\te{in}} (R,\theta)
%   &=
%   \psi_{\te{out}}(R,\theta)
%   \quad
%   \te{for}
%   \quad
%     0 \leq \theta \leq \pi,    
%   \\[0.5em]
%   \label{eq:heat_flux_cont_uncapped_psi}
%   \partial_r \psi_{\te{in}} (R,\theta)
%   &=
%   \partial_r \psi_{\te{out}}(R,\theta)
%   \quad
%   \te{for}
%   \quad
%     0 \leq \theta \leq \pi/2,
%   \\[0.5em]
%   \label{eq:thick-cap_psi}
%   \psi(R,\theta)
%   &=
%     \ln
%     \left(
%     1
%     +
%     \frac{\Delta T}{T_0}
%     \right)
%   \quad
%   \te{for}
%   \quad
%   \pi/2 \leq \theta \leq \pi.
% \end{align}
Hence, the solution for $\psi(r,\theta)$ equals the one given in Eq.~\eqref{eq:Tfield_thick_cap_const_kappa} upon replacing $T_0 \to 0$ and
\(
\Delta T
\to
\ln(1 + \Delta T/T_0)
\).
The resulting temperature field eventually takes the form
\begin{equation}
  \label{eq:T_thick_cap_kappa=1/T}
  T_\kappa(r,\theta)
  =
  T_0
  \left(
    1 + \frac{\Delta T}{T_0}
  \right)^{
    \sum\limits_{n=0}^\infty
    B_n P_n(\cos \theta)
    \left(
      \frac{R}{r}
    \right)^{n+1}
  }.
\end{equation}
As detailed in the supplemental material, an analytic expression for the radially averaged temperature field $\langle T_\kappa\rangle_\theta (r)$ can be calculated when truncating the infinite series in the exponent of Eq.~\eqref{eq:T_thick_cap_kappa=1/T} after $n=1$ (dipole) or $n=2$ (quadrupole).
The solid curve in Fig.~\ref{fig:Tprofile} (A) corresponds to the averaged dipole approximation, which represents the simulation data  very well, also close to the particle surface. The quadrupole approximation is practically indistinguishable from it and is therefore not depicted in Fig.~\ref{fig:Tprofile} (A).
Remarkably, also the temperature field \cite{Chakraborty2011generalEinsteinRel}
\begin{equation}
  \label{eq:T_hom}
  T_{\te{hom}}(r)
  =
  T_0 
  \left(
    1
    +
    \frac{ \ev{\Delta T}_\theta } {T_0}
  \right)^{R/r}  
\end{equation}
of an isotropic particle homogeneously heated up by $\ev{\Delta T}_\theta$  delivers great fits to the data. Here,  $\ev{\Delta T}_\theta$ served as a fit parameter. The corresponding temperature profiles (dashed lines) are almost indistinguishable from the solid curves.
This indicates that, close to the particle surface, the characteristic decay of $\ev{T}_\theta(r)$ is significantly influenced by the temperature dependence of the heat conductivity $\kappa(T)$, whereas heterogeneities in the particle-fluid interactions can be subsumed into a single (fit-)parameter $\ev{\Delta T}_\theta$.
In order to resolve the angle dependence of the temperature profile, we have to resort back to Eq.~\eqref{eq:T_thick_cap_kappa=1/T}.
Using the obtained optimal fit parameters for $T_0$ and $\Delta T$, we numerically determined the angle-resolved temperature profiles $\ev{T_\kappa}_r(\theta)$. They improve the quantitative agreement with the simulation data, as the solid curves in Fig.~\ref{fig:Tprofile} (B) and (C) indicate.

The measured temperature profiles in Fig.~\ref{fig:Tprofile} (B) also show that for $\theta \sim \pi/2$, the metal cap is not well described by an isotherm. Thus, our theoretical prediction slightly overestimates the temperature profile in those regions.
Also, close to the pole of the particle's uncoated hemisphere ($\theta \sim 0$) the measurements indicate a slight temperature increase -- an effect that fades further away from the particle surface as Fig.~\ref{fig:Tprofile} (C) shows.
The investigation of this effect bears potential for future studies as it might indicate a feedback of the particle's hydrodynamic flow field \cite{Bickel2013flowPatternJP} onto the temperature profile. Further improvement of the fits might be obtained by taking the numerically observed Kapitza heat restistance into account.
% A coupling between the temperature field and the stagnation point of the flowfield induced by the swimmer might deliver an explanation for this effect and bears potential for future investigation.

% \begin{figure}[tb!]
%   \centering
%   \includegraphics[width=0.7\columnwidth]{figures/stagnation_point_flowfield_puller.pdf}
%   \caption{Puller flowfield induced by a Janus particle of radius $R$ propelling into the direction of the upper right corner.  The red dot indicates the stagnation point.}
%   \label{fig:stagnation_point}
% \end{figure}

Having justified the notion of a local molecular temperature, we next exploit the aforementioned Brownian time scale separation in order to calculate the effective nonequilibrium temperatures $T_{\HBM}$ that characterize the Janus particle's overdamped hot Brownian motion.

\section{Hot Brownian motion}
\label{sec:HBM}
A hot nanoswimmer is inevitably subject to Brownian motion which randomizes the path of the particle in both position and orientation.
In the classical Langevin picture of equilibrium Brownian motion, the Sutherland-Einstein relation
\begin{equation}
  \label{eq:FDT_classic}
  D = \kBT_0 / \zeta
\end{equation}
for the particle diffusivity $D$ guarantees that the stochastic forces driving the Brownian particle balance the losses by the friction $- \zeta \vv V$, with friction coefficient $\zeta$ and velocity $\vv V$, such as to maintain the Gibbs equilibrium at the temperature $T_0$. 
Equation \eqref{eq:FDT_classic} links the atomistic world, represented by the Boltzmann constant $\kB$, to mesoscopic transport coefficients, $D$ and $\zeta$.
The existence of such a fluctuation-dissipation relation is often taken for granted even when there are temperature gradients present in the solvent, as is the case for heated nanoparticles.
In this situation, however, the stochastic force on the particle must be evaluated as the superposition of the thermal fluctuations within the whole solvent.
In the Markov limit, \ie, on time scales where the particle's momentum and hydrodynamic modes have fully relaxed and Brownian fluctuations are effectively diffusive  \cite{Falasco2016exactSymmHotSwimmer}, generalized overdamped Langevin equations of the form
\begin{equation}
  \label{eq:generalized_Langevin}
  0
  =
  \begin{pmatrix}
    M & 0 \\ 0 & \vv I
  \end{pmatrix}
  \cdot 
  \begin{pmatrix}
    \dot{\vv V} \\ \dot{\vv \Omega}
  \end{pmatrix}
  = - \vv Z \cdot
  \begin{pmatrix}
    \dot{\vv V} \\ \dot{\vv \Omega}
  \end{pmatrix}
  + \vv \xi(t)
  +
  \begin{pmatrix}
    \vv F_{\te{ext}} \\ \vv T_{\te{ext}}
  \end{pmatrix}
\end{equation}
are found to hold \cite{Falasco2016ninisothermFluctHydr,Falasco2014effTempHBM}.
Here, $M, \, \vv I$ denote the particle's mass and tensor of inertia, $\vv V, \, \vv \Omega$ its translational and rotational velocity, $\vv Z$ the $6 \times 6$ friction tensor, $\vv \xi(t)$ Gaussian white noise, and $\vv F_{\te{ext}}, \, \vv T_{\te{ext}}$ the external force and torque, respectively.
The equations of motion are complemented by the specification of the noise strength
\begin{align}
  \label{eq:zero_mean_HBM}
  \langle \vv \xi(t) \rangle
  &=
    \vv 0,
  \\[0.5em]
  \label{eq:FDR_HBM}  
  \langle \vv \xi^\mu(t) \vv \xi^\nu(t') \rangle 
  &\propto
    T_\HBM^{\mu \nu} Z_\HBM^{\mu \nu} \delta(t-t'),
\end{align}
with effective temperatures $T_\HBM^{\mu \nu}$ and friction coefficients $Z_\HBM^{\mu \nu}$, respectively.
The superscript $\mu \nu$ indicates that for non-isothermal Brownian motion, different degrees of freedom (\eg, translation, rotation, or both coupled) sense distinct effective temperatures \cite{Falasco2014effTempHBM} rendering $T_\HBM$, in general, a tensorial quantity.
Using the framework of fluctuating hydrodynamics, the effective temperatures $T_\HBM^{\mu \nu}$ turn out to be given by a weighted spatial average of the local solvent temperature field $T(\vv r)$ \cite{Falasco2016ninisothermFluctHydr}:
\begin{equation}
  \label{eq:T_HBM}
  T_\HBM^{\mu \nu} =
  \frac{
    \int \df \vv r \, T(\vv r) \phi^{\mu \nu}(\vv r)
  }{
    \int \df \vv r \, \phi^{\mu \nu}(\vv r)
  }.
\end{equation}
The weight function
\begin{equation}
  \label{eq:phi_fct}
  \phi^{\mu \nu}(\vv r) 
  \equiv
  \eta(\vv r)
  \sum\limits_{i,j}
  \left[
    \partial_i u_j^\mu(\vv r)
    \partial_i u_j^\nu(\vv r)
    +
    \partial_i u_j^\nu(\vv r)
    \partial_j u_i^\mu(\vv r)
  \right]
\end{equation}
is the (excess) viscous dissipation function induced by the velocity fields $\vv u^{\mu, \nu}$ pertaining to the considered motion and $\eta$  is the dynamic viscosity of the fluid. Note that $T_\HBM^{\mu \nu}$ is therefore a tensorial quantity.

We stress the fact that the theory of hot Brownian motion connects the particle's enhanced thermal fluctuations with the associated energy dissipation into the ambient fluid. Therefore, the dissipation function $\phi^{\mu\nu}$ defined in Eq.~\eqref{eq:phi_fct} must \emph{not} include contributions due to the particle's active swimming, even though the latter typically exceeds the former considerably.

In the following section, we use Eq.~\eqref{eq:T_HBM} to estimate effective temperatures characterizing the rotational and translational hot Brownian motion of a Janus sphere.

\section{Estimating $T_\HBM$ for a Janus sphere}
\label{sec:THBM_JanusSphere}
Since a generally temperature dependent viscosity $\eta(T)$ renders the calculation of effective temperatures rather complicated, we pursue a similar approach as in~\cite{Rings2011theoryOfHBM,Rings2012rotHBM}, where a first estimate for $T_\HBM^{\mu \nu}$ could be obtained by employing a temperature independent viscosity $\eta \equiv \eta_0$.  In the case of a homogeneously heated particle \cite{Rings2011theoryOfHBM,Rings2012rotHBM} with a surface temperature increment $\Delta T$, this lead to the first order term in $T_\HBM(\Delta T)$.
Following the same route, we calculate the effective temperatures $T_\HBM^{\theta_{\parallel/\perp}}$ and $T_\HBM^{x_{\parallel/\perp}}$
for a Janus sphere. The superscripts represent motion types considered in this article, namely
\begin{itemize}
\item $\theta_{\parallel/\perp}$: rotation about/perpendicular to the particle's symmetry axis,
\item $x_{\parallel/\perp}$: translation along/transverse to the symmetry axis.
\end{itemize}
The effective temperature $T_\HBM^m$ corresponding to the considered motion type
\(
m
\in
\{\theta_{\parallel/\perp}, x_{\parallel/\perp}\}
\)
is given by
\begin{equation}
  \label{eq:app_T_HBM}
  T_\HBM^m =
  \frac{
    \int \df \vv r \, T(\vv r) \phi^m(\vv r)
  }{
    \int \df \vv r \, \phi^m(\vv r)
  }.
\end{equation}
Note that superpositions of the motion types listed above generally sense yet different effective temperatures, \eg, $T_\HBM^{x_\parallel,\theta_\perp}$. Here, we only consider the elementary degrees of freedom.

The temperature field around a Janus sphere of radius $R$ solves the heat conduction equation \eqref{eq:heat_equation_general}. Assuming constant viscosity and heat conductivity $\kappa$, the solution can be expanded in terms of Legendre polynomials $P_n$ as
\begin{equation}
  \label{eq:app_Tfield_general}
  T(r,\theta)
  =
  T_0
  +
  \sum\limits_{n=0}^\infty
  \mathcal T_n P_n(\cos\theta)
  \left(
    \frac{R}{r}
  \right)^{n+1},
\end{equation}
with the ambient fluid temperature $T_0$. The expansion coefficients $\mathcal T_n$ are determined by boundary conditions at the particle surface, which we do not further specify here.
Analogous to the calculation in Eqs.~\eqref{eq:mean_T_of_r_first}-\eqref{eq:mean_T_of_r}, the average surface temperature increment is given by
\begin{align}
  \label{eq:mean_surf_temp_Thbm}
  \ev{\Delta T}_\theta
  =
  \mathcal T_0.
\end{align}
Since the coefficient $\mathcal T_0$ pertains to an angle independent $1/r$-decay in the temperature field \eqref{eq:app_Tfield_general}, it represents a spherical particle homogeneously heated to $T_0+\ev{\Delta T}_\theta$. Higher coefficients $\mathcal T_{n>0}$ characterize  anisotropies of the temperature field.

We now turn to the calculation of the effective temperature $T_\HBM^{x_\parallel}$ via Eq.~\eqref{eq:app_T_HBM}, where we first consider the Janus particle's translation along its symmetry axis. Introducing the abbreviations $s \equiv \sin\theta$ and $c \equiv \cos\theta$, the corresponding viscous dissipation function for a sphere of radius $R$ at translation speed $V$ within an infinite homogeneous system with constant viscosity reads \cite{Rings2011theoryOfHBM}
\begin{equation}
  \label{eq:app_diss_fct_trHBM}
  \phi^x(r,\theta)
  =
  \eta_0
  \left[
    f_s(r) s^2 + f_c(r)c^2
  \right],
\end{equation}
where we introduced
\begin{align}
  \label{eq:fs}
  f_s(r)
  &\equiv
    \frac{9}{r^8}
    K_2^2,
  \\
  \label{eq:fc}
  f_c(r)
  &\equiv
    \frac{3}{r^8}
    \left(
    K_1 r^2 +  3K_2
    \right)^2,
\end{align}
with the constant coefficients $K_1=-3 V R / 2$ and $K_2 = V R^3/2$.  As the dissipation function $\phi^x$ is composed of terms proportional to $s^2$ and $c^2 = 1-s^2$, the product of $T(r,\theta)$ and $\phi^x(r,\theta)$ in the enumerator of Eq.~\eqref{eq:app_T_HBM} yields integrals over the polar angle of the kind
\(
\int_{-1}^1 \df c ~ s^2 P_n(c).
\)
By virtue of the relation $s^2 = 2[P_0(c) - P_2(c)]/3$ and Eq.~\eqref{eq:Pn_orth}, this integral simplifies to
\begin{equation}
  \label{eq:APP_Legendre_int}
  \int_{-1}^1 \df c \, 
  s^2 P_n(c)
  =
  \frac23
  \left(
    \frac{2 \delta_{0n}}{2n+1}
    -
    \frac{2 \delta_{2n}}{2n+1}
  \right).
\end{equation}
Therefore, only the coefficients $\mathcal T_0$ and $\mathcal T_2$ from the infinite series \eqref{eq:app_Tfield_general} contribute to $T_\HBM^{x_\parallel}$, while all other terms vanish by symmetry upon averaging, which actually holds for all cases considered in this article. Explicit calculation of $T_\HBM^{x_\parallel}$ via Eq.~\eqref{eq:app_T_HBM} gives for the non-trivial part of the numerator:
\begin{align}
  &2 \mathcal T_0
    \int\limits_R^\infty \df r ~
    r^2 f_s(r)
    \frac{R}{r}
  \\
  &+
    \frac23 \mathcal T_0
    \int\limits_R^\infty \df r ~
    r^2
    \left[
    f_c(r) - f_s(r)
    \right]
    \frac{R}{r}
  \\
  \label{eq:Thbm_x_vanishing_int}  
  &+
  \frac{4}{15} \mathcal T_2
  \int\limits_R^\infty \df r ~
    r^2
    \left[
    f_c(r) - f_s(r)
    \right]
    \left(
    \frac{R}{r}
    \right)^3
  \\[0.5em]
  &=
    \frac34 \mathcal T_0 R V^2
    +
    \frac12 \mathcal T_0 R V^2
    +
    0
  \\
  \label{eq:enumerator_Tbhm_tr}  
  &=
    \frac54 \mathcal T_0 R V^2.
\end{align}
The denominator analogously gives $3 R V^2$.
With Eq.~\eqref{eq:mean_surf_temp_Thbm}, we finally obtain
\begin{equation}
  \label{eq:Thbm_tr_par}
  T_\HBM^{x_\parallel}
  =
  T_0
  +
  \frac{5}{12}
  \ev{\Delta T}_\theta.
\end{equation}
Hence, for translation along its symmetry axis, the Janus sphere's hot Brownian motion is identical to that of a sphere homogeneously heated by $\ev{\Delta T}_\theta$ \cite{Rings2011theoryOfHBM}.  The vanishing integral in line \eqref{eq:Thbm_x_vanishing_int} reveals that there is no coupling between the dissipation function $\phi^x$ and angular variations in the temperature field represented by $\mathcal T_2$.  Therefore, no correction term accompanies the factor $5/12$ in $T_\HBM^{x_\parallel}$. This result could have been anticipated as the authors of Ref.~\cite{Falasco2016ninisothermFluctHydr} showed that a linear temperature field implies no corrections to the standard Langevin description of Brownian motion when the viscosity is assumed to be constant.

As anticipated on the same grounds and explicitly shown in the supplemental material,  a similar calculation with a simple coordinate transformation leads to
\begin{align}
  \label{eq:Thbm_tr_perp}
  T_\HBM^{x_\perp}
  =
  T_0
  +
  \frac{5}{12}  
  \ev{\Delta T}_\theta,
\end{align}
for the particle's translation perpendicular to its symmetry axis.
Thus, also for transverse motion the corresponding effective temperatures are exactly given by those of a sphere homogeneously heated up by $\ev{\Delta T}_\theta$.

We now turn to the particle's rotational degrees of freedom starting with the calculation of $T_\HBM^{\theta_\parallel}$ for rotation about its symmetry axis. The corresponding viscous dissipation function for a rotating sphere reads \cite{Rings2012rotHBM}
\begin{equation}
  \label{eq:app_diss_fct_rotHBM}
  \phi^\theta(r,\theta) 
  \propto
  r^{-6} \sin^2 \theta.
\end{equation}
 Using Eq.~\eqref{eq:APP_Legendre_int}, the non-trivial part of the numerator of Eq.~\eqref{eq:app_T_HBM} evaluates to
\begin{align}
  \label{eq:1}
    &\frac23 \sum\limits_{n=0}^\infty
    \int\limits_R^\infty \df r \int\limits_{-1}^1 \df c \,
    P_n(c) \frac{1}{r^4}
    \left(
      \frac{R}{r}
    \right)^{n+1}
    [P_0(c) - P_2(c)]
    \\[0.5em]
    &= \frac23
    \left[
      2 \mathcal T_0 \int_R^\infty \df r \, \frac{R}{r^5}
      -\frac25 \mathcal T_2 \int_R^\infty \df r \, \frac{R^3}{r^7}
    \right]
  \\[0.5em]
  \label{eq:APP_numerator_T_HBM}  
    &= \frac{4}{9R^3}
    \left[
      \frac34 \mathcal T_0 - \frac{1}{10} \mathcal T_2
    \right],  
\end{align}
and likewise the denominator to
\begin{equation}
  \label{eq:APP_denominator_T_HBM}
  \frac43 \int_R^\infty \df r \, r^{-4} = \frac{4}{9 R^3}.
\end{equation}
Using $\mathcal T_0 = \ev{\Delta T}_\theta$ finally yields
\begin{equation}
  \label{eq:Thbm_rot_par}
  T_\HBM^{\theta_\parallel}
  =
  T_0
  +
  \ev{\Delta T}_\theta
  \left(
    \frac34
    -
    \frac{1}{10}
    \frac{\mathcal T_2}{\ev{\Delta T}_\theta}
  \right).
\end{equation}
In contrast to our results for translation, Eqs.~\eqref{eq:Thbm_tr_par} and \eqref{eq:Thbm_tr_perp}, we find that $T_\HBM^{\theta_\parallel} - T_0$ is composed of two parts: (i) the contribution $3 \ev{\Delta T}_\theta/4$ corresponding to an isotropic sphere homogeneously heated up by $\ev{\Delta T}_\theta$ \cite{Rings2012rotHBM} and (ii) a correction term proportional to $\mathcal T_2$. The latter stems from a non-vanishing coupling between the $\theta$-dependence of the temperature field \eqref{eq:app_Tfield_general} and the dissipation function \eqref{eq:app_diss_fct_rotHBM}. For most practical cases, the correction term is negligible with respect to $3/4$ since, typically, $\mathcal T_2 < \mathcal T_0 = \ev{\Delta T}_\theta$. For instance, in the thick-cap limit, for which the temperature field is explicitly given by Eqs.~\eqref{eq:Tfield_thick_cap_const_kappa} and \eqref{eq:Legnde_Coeffs}, one has
\(
\mathcal T_2 / (10 \mathcal T_0) \approx -0.03.
\)

A simple coordinate transformation (see supplemental material) and similar calculations as presented above yield
\begin{equation}
  \label{eq:Thbm_rot_perp}
  T_\HBM^{\theta_\perp}
  =
    T_0
    +
    \ev{\Delta T}_\theta
    \left(
    \frac34
    +
    \frac{1}{20}
    \frac{\mathcal T_2}{\ev{\Delta T}_\theta}
    \right),
\end{equation}
for rotation perpendicular to the particle's symmetry axis.
In this case, the correction term is only half in magnitude and has opposite sign as compared to the one in Eq.~\eqref{eq:Thbm_rot_par}.  This stems from the fact that the symmetry axis of the particle, and thus of the temperature profile, does not coincide with the rotation axis, thus leading to a distinct coupling between the temperature field and the dissipation function.

\begin{figure}[tb!]
  \centering
\includegraphics[width=0.5\columnwidth]{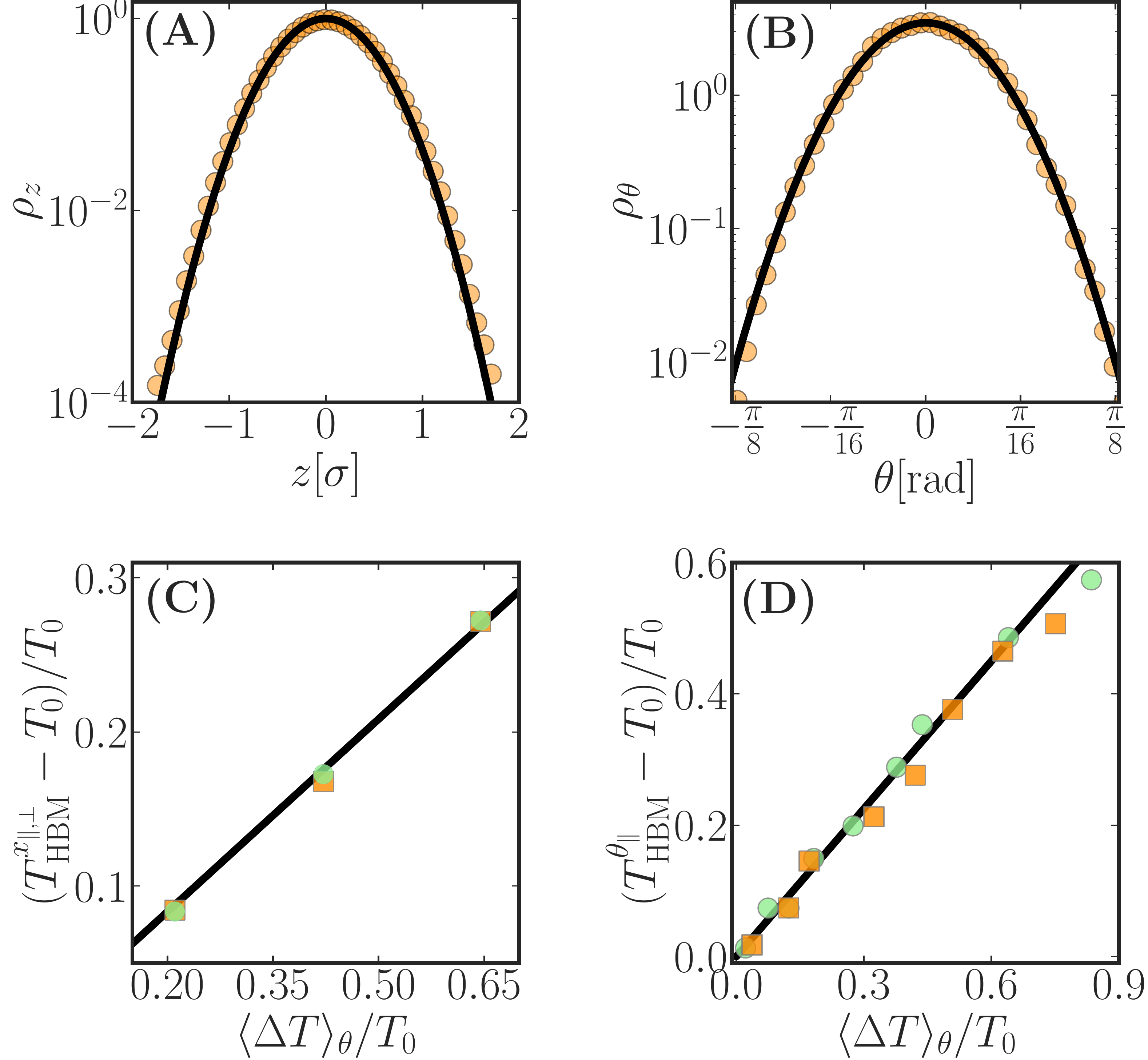}  
\caption{%
  Upper panels: Histograms of harmonically confined position \textbf{(A)} and orientation \textbf{(B)} of the Janus particle heated to $T_{\rm P} = \SI{1.0}{\epsilon/\kB}$ in \textbf{(A)} and $T_{\rm P} = \SI{0.9}{\epsilon/\kB}$ in \textbf{(B)}.
    Histograms were fitted by Gaussian profiles (black curves) to extract the corresponding effective temperature $T_\HBM$.
    \textbf{(C)} numerically measured $T_{\rm HBM}^{x_{\parallel,\perp}}$ for translation along/transverse  to the symmetry axis (red squares/green circles) against the theory (black line) from Eq.~\eqref{eq:Thbm_tr_par},\eqref{eq:Thbm_tr_perp}.
\textbf{(D)} $T_\HBM^{\theta_\parallel}$ for rotation around the particle's symmetry axis for a heated Janus particle (orange squares), and $T_\HBM^\theta$ for an isotropic, homogeneously heated Brownian particle \cite{Rings2012rotHBM} (green circles) versus theory (black curve) from Eq.~\eqref{eq:Thbm_rot_par}, neglecting the small correction term (which would hardly be discernible, anyway). In all simulations, the wetting parameters of the interaction potential were set $c_{\te{ps}} = 1$ and $c_{\te{gs}} = 2$.
\label{fig:Thbm_xDist_thetaDist}
  }
\end{figure}

In order to test the theory, we measured the effective temperatures using MD computer simulations as described in Sec.~\ref{sec:an-atom-impl-1}.  
We therefor confined the Janus sphere to an angular or spatial harmonic potential parallel or perpendicular to its symmetry axis.
Figures.~\ref{fig:Thbm_xDist_thetaDist} (A) and (B) illustrate the respective distributions of the Janus particle's position $z$ and orientation $\theta$ relative to its symmetry axis for distinct heating temperatures.
From the variances of the effectively Gaussian distributions, we extracted the effective temperatures for translation and rotation, respectively. The corresponding average surface temperature increments $\langle \Delta T \rangle_\theta$ were obtained by fitting the angle averaged temperature profiles via Eq.~\eqref{eq:T_hom} (cf.~Fig.~\ref{fig:Tprofile}).
As Fig.~\ref{fig:Thbm_xDist_thetaDist} (C) and (D) show, the measured effective temperatures are nicely described by our theory.
We also compared our results to the effective temperatures for a homogeneously heated Brownian particle in Fig.~\ref{fig:Thbm_xDist_thetaDist} (D).

Having seen that the coarse-grained non-equilibrium hydrodynamic description works well on the single-particle level, we now address the task of further coarse graining a suspension of hot microswimmers to an effective homogeneous complex fluid.

\section{Complex fluid homogenisation}
\label{sec:CoarseGraining}

One is often interested in the collective (thermo)dynamical properties of an assembly of colloids and their embedding solvent, rather than in the motion of a single unit. Particle-based descriptions are impractical to inspect the behavior of such a complex fluid and one therefore often seeks a more versatile continuum approach, which allows one to leapfrog, in an efficient way, over the diverse time and length scales of its various constituents.

To this aim, we study non-isothermal fluctuating hydrodynamic equations of a fluid with suspended colloids, which are recast in terms of dynamical equations for coarse-grained volume elements. Surprisingly, a non-local frequency-dependent temperature appears due to the presence of the dispersed particles, which characterizes the intensity of their thermal fluctuations.
Consider an incompressible solvent of density $\varrho$ with velocity field $\vv u$ described by the linearized fluctuating hydrodynamic equations~\cite{Falasco2014effTempHBM}
\begin{equation}
  \label{eq:Stokes_Equ}
  \varrho \partial_t \vv u(\vv r',t)
  =
  \vv \nabla \cdot \vv \sigma(\vv r',t)
  +
  \vv \nabla \cdot \vv \tau(\vv r',t),
\end{equation}
and $N$ suspended colloids coupled to the fluid velocity via no-slip boundary conditions.
Here, $\vv \sigma$ denotes the total stress tensor and $\vv \tau$ is a zero-mean Gaussian random stress tensor with delta correlations in space and time.
The molecular temperature is prescribed by the heat equations.
For later convenience, we assume that the flow field $\vv u$ is defined also inside the colloids, where it equals identically the colloid velocity $\vv V_i$, namely
\begin{equation}
  \label{eq:colloid_Velocity}
  \vv u(\vv r',t) = \vv V_i, 
  ~ \te{if} ~ 
  \MYabs{ \vv r' - \vv X_i' } < R.
\end{equation}
The colloids are idealized as spheres with radius $R$ and mass $M$, and their positions are denoted by $\vv X_i$.

The coarse-graining procedure proceeds as follows:
\begin{enumerate}
\item We divide the system into mesoscopic volumes $\mathcal V(\vv r)$, with edge length $l$, located at position $\vv r$.
Old and new coordinates, $\vv r'$ and $\vv r$ respectively, are related by the scaling $\vv r = \vv r' /l$, where $l$ defines the coarse-graining length scale, much larger than the colloidal radius $R$.
\item We define the coarse-grained velocity of the complex fluid $\vv U (\vv r, t)$ as the spatial average over
the coarse-grained volume $\mathcal V(\vv r)$
\begin{equation}
  \label{eq:average_Velocity}
  \vv U(\vv r,t) = \frac 1 {l^3}
  \int\limits_{\mathcal V(\vv r)} \df \vv r' \,
  \vv u(\vv r',t).
\end{equation}
\item In Eq.~\eqref{eq:average_Velocity} the integration volume $\mathcal V(\vv r)$ is split into the volume occupied by the solvent $\mathcal V_{\te s} (\vv r)$ and that occupied by the solute particles $\mathcal V_{\te p}(\vv r)$.
  Introducing the local particle volume fraction $\phi(\vv r) \equiv \MYabs{ \mathcal V_{\te p}(\vv r) } / \MYabs{ \mathcal V(\vv r) }$, we obtain
  \begin{equation}
    \label{eq:average_Velocity_2}
    \vv U(\vv r,t) = \frac 1 {l^3}
    \int\limits_{\mathcal V_{\te s}(\vv r)} \df \vv r' \,
    \vv u(\vv r',t)
    +
    \phi(\vv r) \overline{\vv V}(\vv r,t),
  \end{equation}
where we have defined the local average velocity of the colloids
\begin{align}
  \label{eq:local_average_Velocity}
  \overline{\vv V}(\vv r,t)
  &\equiv
  \frac 1 {N(\vv r)}
  \sum\limits_{i \in \mathcal V(\vv r)} \vv V_i(t),
  \\
  \label{eq:local_particle_number}  
  N(\vv r)
  &\equiv
  \sum\limits_{i \in \mathcal V(\vv r)} 1.
\end{align}
\item We take the time derivative of Eq.~\eqref{eq:average_Velocity_2},
  \begin{align}
    \label{eq:averageVelocity_by_dt}
    &\partial_t \vv U(\vv r,t)
    =
      \frac 1 {l^3}
    \int\limits_{\mathcal V_{\te s}(\vv r)} \df \vv r' \,
    \partial_t \vv u(\vv r',t)
    +
    \phi(\vv r) \partial_t \overline{\vv V}(\vv r,t).
  \end{align}
The first summand is rewritten, using Eq.~\eqref{eq:Stokes_Equ}
\begin{align}
    &\frac 1 {l^3}
    \int\limits_{\mathcal V_{\te s}(\vv r)} \df \vv r' \,
    \partial_t \vv u(\vv r',t)
    \\[0.5em]
    &= 
    \frac 1 {\varrho l^3}
    \int\limits_{\mathcal V_{\te s}(\vv r)} \df \vv r' \,
    \left(
      \vv \nabla \cdot \vv \sigma(\vv r',t)
      +
      \vv \nabla \cdot \vv \tau(\vv r',t)
    \right)
    \\[0.5em]
    &=
    \frac 1 \varrho
    \left(
      \vv \nabla \cdot \vv \sigma_\nu(\vv r,t)
      +
      \vv \nabla \cdot \vv \tau_\nu(\vv r,t)
    \right),    
\end{align} 
where $\vv \sigma_\nu$ and $\vv \tau_\nu$ are the coarse-grained stress tensors.
\end{enumerate}

It is not hard to convince oneself that the random tensor $\vv \tau_\nu$ is still a zero-mean Gaussian noise with delta correlations in space and time.
In fact, using the divergence theorem the mean is
\begin{equation}
  \label{eq:mean_tau}
    \int\limits_{\partial \mathcal V_{\te s}(\vv r)} \df^2 r' \,
    \vv n' 
    \cdot
    \ev{ \vv \tau(\vv r',t) }
    =
    \vv 0,
  \end{equation}
and the correlation function becomes
\begin{equation}
  \begin{split}
    &\int\limits_{\partial \mathcal V_{\te s}(\vv r_1)} \df^2 r'
    \int\limits_{\partial \mathcal V_{\te s}(\vv r_2)} \df^2 r'' \,
    \vv n' \vv n'' 
    \cdot
    \ev{%
      \vv \tau(\vv r',t) \vv \tau(\vv r'',t)
    } 
    \\[0.5em]
    =
    2&\int\limits_{\partial \mathcal V_{\te s}(\vv r_1)} \df^2 r'
    \int\limits_{\partial \mathcal V_{\te s}(\vv r_2)} \df^2 r'' \,
    \vv n' \vv n'' \eta(\vv r) T(\vv r) 
    \delta(\vv r' - \vv r'') \delta(t - t'),    
  \end{split}
  \label{eq:correlation_tau}
\end{equation}
which is nonzero only if $\vv r_1 = \vv r_2$ and is proportional to $\delta(t-t')$. Here, $\vv n'$ and $\vv n''$ denotes the inner normal vectors along the surface $\partial \mathcal V_\te{s}(\vv r_{1,2})$ of the respective volume element.
The second summand of equation \eqref{eq:averageVelocity_by_dt} is simply
\begin{align}
  \label{eq:dU_by_dt_second}
  \frac{\phi(\vv r)}{N(\vv r)} 
    \sum\limits_{\mu \in \mathcal V(\vv r)}
    \dot{\vv V}_\mu(t)
  =\frac{\phi(\vv r)}{N(\vv r) M}
    \sum\limits_{\mu \in \mathcal V(\vv r)}
  \vv v_{\te{tp}} 
    -
    \int_{-\infty}^t \df t' \,
   \vv Z(\vv X_\mu, t-t') \cdot \vv V_\mu(t')
    +
    \vv \xi_\mu(\vv X_\mu,t),
\end{align}
with the propulsion speed $\vv v_{\te{tp}} $ induced by the local temperature field, the time-dependent friction kernel $\vv Z(\vv X_\mu,t)$ and the unbiased white Gaussian noise process $\vv \xi_\mu(\vv X_\mu,t)$.

Therefore, the average acceleration is given by a sum of generalized Langevin equations of the type previously derived for hot Brownian motion~\cite{Falasco2014effTempHBM,Falasco2016ninisothermFluctHydr}.
We use the result obtained for a generic (but stationary) temperature field $T(\vv r')$, here generated by the colloids sitting in their average positions. The noise spectrum reads
\begin{equation}
  \begin{split}
    &\langle \vv \xi_\mu^i(\vv
    X'_\mu,\omega) \vv \xi_\mu^j(\vv
    X'_\mu,\omega) \rangle
    \propto
    \mathcal T^{\mu \nu}(\vv X_\mu', \omega) Z^{\mu \nu}(\vv X_\mu',\omega) 
  \end{split}
\end{equation}
 and displays a tensorial frequency-dependent temperature --- generally distinct from $T(\vv r')$ --- 
 \begin{equation}
  \label{eq:noise_temperature}
  \mathcal T^{\mu \nu}(\vv X_\mu', \omega)
  \equiv
  \frac{
    \int\limits_V \df^3 r' \,
    \phi^{\mu \nu}(\vv r', \omega) T(\vv X_\mu' + \vv r')
  }{
    \int\limits_V \df^3 r' \,
    \phi^{\mu \nu}(\vv r', \omega)
  },
\end{equation}
whose zero-frequency limit should be compared with Eq.~\eqref{eq:T_HBM}. The weight function $ \phi^{\mu \nu}(\vv r', \omega)$ denotes the (excess) viscous dissipation function associated to the unsteady thermal motion of a colloid.
For this to apply to all particles in the suspension, the latter has to be sufficiently dilute so that hydrodynamic and temperature-mediated interactions between the colloids are negligible.
Under this assumption, also thermal noises acting on different colloids can be taken to be uncorrelated.

This analysis suggests that a non-trivial coarse-grained noise temperature arises through the presence of ``slow'' degrees of freedom.
These are, in a hydrodynamic description, coupled to the fast ones via boundary conditions and thus are subjected to long-range forces, in contrast to the local, Markovian thermal stresses acting on the fluid elements.

\section{Conclusion}
\label{sec:conclusion}

We have performed microscopically resolved  molecular dynamics simulations of a single hot Janus swimmer immersed in a Lennard-Jones fluid. We locally measured the inhomogeneous and anisotropic temperature profile induced in the solvent and compared it against analytic expressions basing on the heat conduction equation. We thereby verify the notion of a molecular temperature at which the surrounding medium locally equilibrates. We then exploited a large Brownian timescale separation in order to address the Janus particle's overdamped hot Brownian motion. In a first-order approximation in the mean temperature increment $\langle \Delta T\rangle_\theta $ of the particle surface, we calculate effective nonequilibrium temperatures $T_\HBM$ for distinct types of motion. Our theoretical predictions nicely agree with measurements of $T_\HBM$ over a wide temperature range.
In the last coarse-graining step, we studied non-isothermal fluctuating hydrodynamic equations of a fluid with such suspended nonisothermal active colloids. The noise spectrum of the coarse-grained fluid is governed by tensorial, local and frequency-dependent effective temperatures that generally differ from the local molecular temperature field in the solvent. They have to separately be taken along in any attempt to coarse grain a suspension of hot particles into a an effectively homogeneous complex fluid.

\section*{Conflict of Interest Statement}

The authors declare that the research was conducted in the absence of any commercial or financial relationships that could be construed as a potential conflict of interest.

\section*{Author Contributions}
The simulations were designed and performed by DC and RP. Both also analyzed the raw data. The theory was done by SA and GF. The manuscript was written by SA and KK.

\section*{Funding}
We acknowledge funding by Deutsche Forschungsgemeinschaft (DFG) via SPP 1726 and KR 3381/6-1, and Universität Leipzig within the program of Open Access Publishing.

\newpage
\appendix

{\huge{APPENDIX}}

\def\thesection{\alph{section}}

\section{Propulsion Speed and Direction}
\label{sec:propulsion-speed}

\begin{figure}[h!]  \centering \includegraphics[width=\columnwidth]{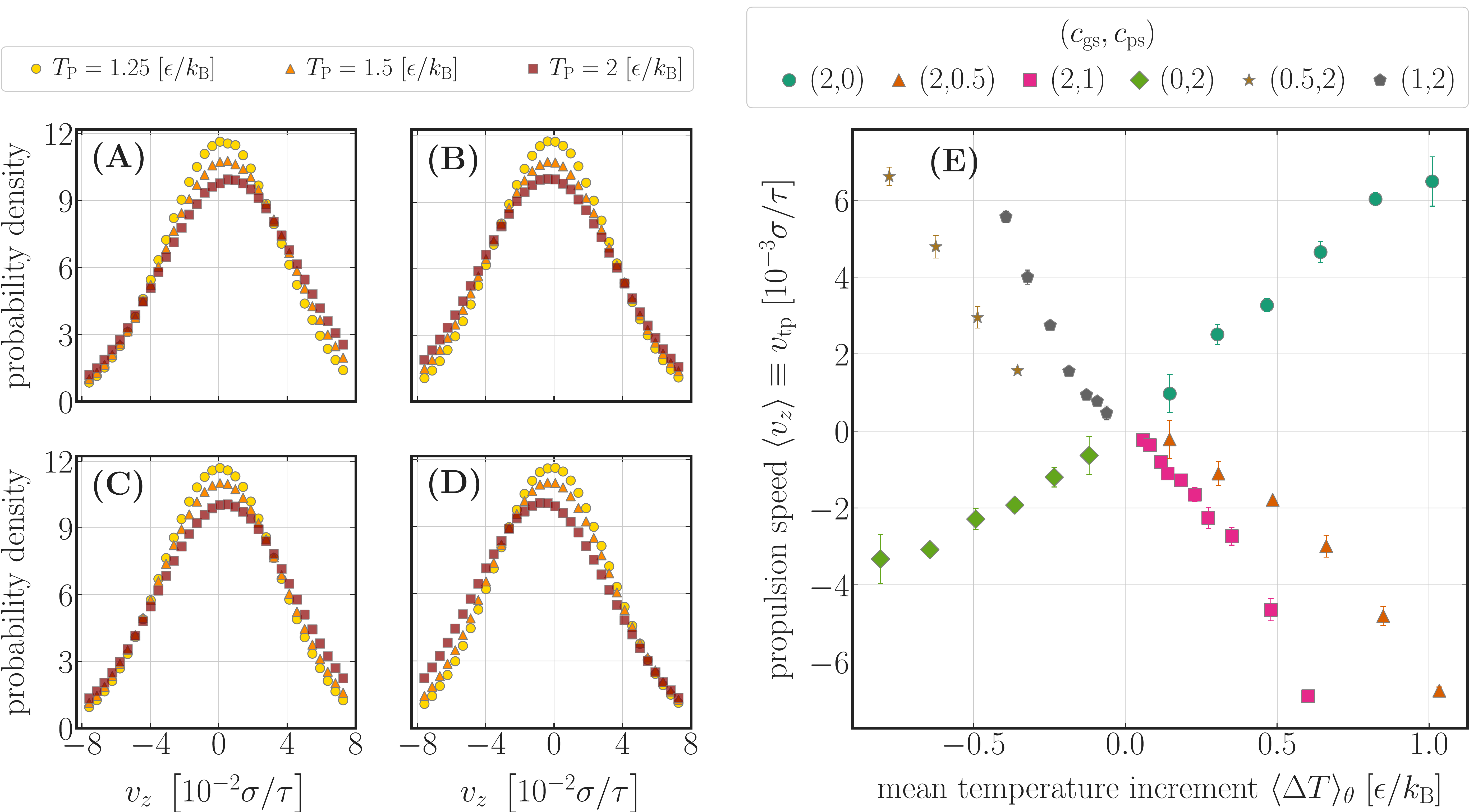}  
  \caption{%
    \textbf{(A)}-\textbf{(D)} Probability distribution of the velocity component $v_z$ along the particle's symmetry axis ($z$-axis) pointing towards the uncapped hemisphere. Distributions are plotted for different heating temperatures and different choices for the wetting parameters [\textbf{(A)}: $c_{\te{ps}} = 2$, $c_{\te{ps}} = 0$, \textbf{(B)}: $c_{\te{ps}} = 0$, $c_{\te{ps}} = 2$, \textbf{(C)}: $c_{\te{ps}} = 1$, $c_{\te{ps}} = 2$, \textbf{(D)}: $c_{\te{ps}} = 2$, $c_{\te{ps}} = 1$].
    The mean value $\langle v_z \rangle$ of the respective distributions is identified with the thermophoretic propulsion speed $v_{\te{tp}}$ of the particle. The latter is plotted in \textbf{(E)} as function of the mean surface temperature increment $\langle \Delta T \rangle_\theta$ of the particle for different choices of the wetting parameters. Here, $\langle \Delta T \rangle_\theta$ was obtained from fitting the measured angle averaged temperature profiles via
    \(
    T(r)
    =
    T_0
    \left(
      1 + \langle \Delta T \rangle_\theta/T_0
    \right)^{R/r}
    \)
    (see main text for details). Note that a reversal in the direction of propulsion is observed when the attractive part of the Lennard-Jones potential is switched off on one hemisphere.
  }
  \label{fig:prop_vel}
\end{figure}

\section{Angle averaged temperature profiles}
\label{sec:angle-aver-temp}

Introducing the auxiliary quantities
\(
\tilde t
\equiv
1 + \Delta T / T_0
\)
and
\(
b_n(r)
\equiv
B_n (R/r)^{n+1}
\),
the temperature profile given in Eq.~(13) of the main text reads
\begin{equation}
  \label{eq:T_kappa}
  T_\kappa(r,c)
  =
  T_0 \te{e}^{
    \ln(\tilde t)
    \sum_{n=0}^\infty b_n(r)P_n(\cos \theta).
  }
\end{equation}
The dipole and quadrupol approximations ($b_{n>1,2} \equiv 0$, respectively) of the above expression read
\begin{align}
  \label{eq:T_kappa_dipole}
  T_\kappa^{(1)}(r,\theta)
  &\equiv
    T_0
    \te{e}^{
    \ln(\tilde t)
    \left[
    b_0 + b_1 P_1(\cos\theta)
    \right]
    },
  \\
  \label{eq:T_kappa_quadrpl}
  T_\kappa^{(2)}(r,\theta)
  &\equiv
    T_0
    \te{e}^{
    \ln(\tilde t)
    \left[
    b_0 + b_1 P_1(\cos\theta) + b_2 P_2(\cos\theta)
    \right]
    }.
\end{align}
Using the abbreviation $c \equiv \cos\theta$, the respective $\theta$-averaged profiles
\begin{equation}
  \label{eq:angle_average_def}
  \langle T_\kappa^{(i)} \rangle_\theta (r)
  \equiv
  \frac12
  \int_{-1}^1 \df c ~
  T_\kappa^{(i)}(r,c),
  \qquad
  i=1,2,
\end{equation}
yield integrals of the form
\begin{align}
  \label{eq:int_dipole}
  \int_{-1}^1 \df c ~
  \te{e}^{D + E c + \delta_{i2} F c^2},
\end{align}
with functions $D(r)$, $E(r)$ and $F(r)$ determined by comparing the exponent in Eq.~\eqref{eq:int_dipole} with Eqs.~\eqref{eq:T_kappa_dipole} and \eqref{eq:T_kappa_quadrpl}, respectively. Here, $\delta_{i,2}=1$ if $i=2$ and zero otherwise. Calculating the integrals, one obtains
\begin{align}
  \label{eq:T_av_dipole}    
  \ev{T_\kappa^{(1)}}_\theta(r)
  &=
    \frac{T_0}{E}
    \sinh(E) \te{e}^D,
  \\[0.5em]
  \label{eq:T_av_quadrpl}  
  \ev{T_\kappa^{(2)}}_\theta(r)
  &=
    \frac{T_0 ~ \te{e}^{D-E+F}}{\sqrt{F}}
  \left[
    \te{e}^{2E}
    \mathcal D
    \left(
    \frac{E+2F}{2 \sqrt{F}}
    \right)
    -
    \mathcal D
    \left(
    \frac{E-2F}{2 \sqrt{F}}
    \right)    
    \right],
\end{align}
with the Dawson integral
\(
\mathcal D(x)
\equiv
\te{e}^{-x^2}
\int_0^x \df y ~ \te{e}^{y^2}.
\)
The profile \eqref{eq:T_av_dipole} is represented by the solid (red) curves in Fig.~2 (A) of the main text. The corresponding profiles according to Eq.~\eqref{eq:T_av_quadrpl} are almost indistinguishable from the former and are thus not depicted.

\section{$T_\HBM$ for transverse motion}
\label{sec:App_T_HBM_Est1}

\begin{figure}[b!]
  \centering  \includegraphics[width=0.4\columnwidth]{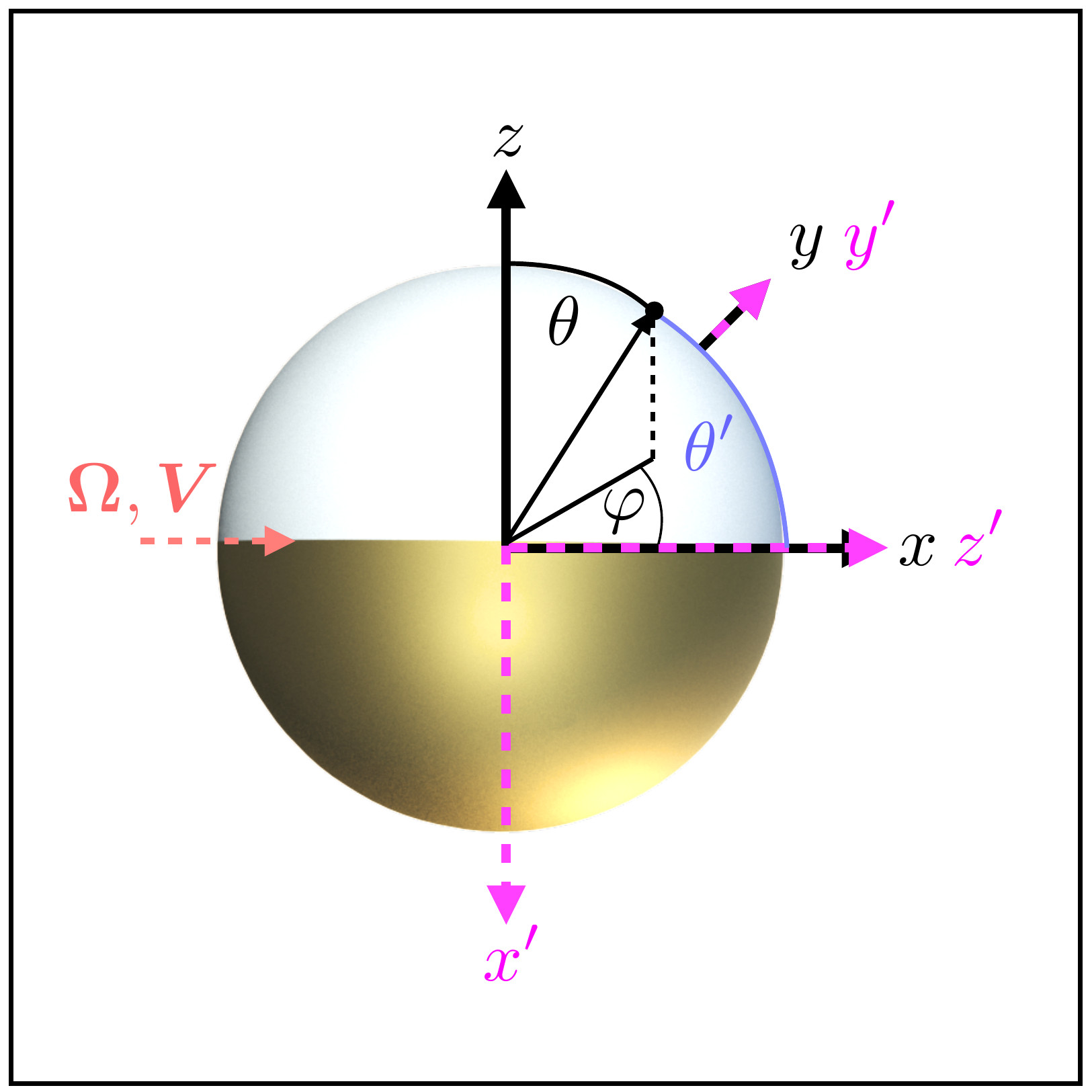}
  \caption{Definition of coordinate systems. Polar angles, $\theta$ and $\theta'$, are measured relative to the respective $z$-axes, whereas azimuthal angles, $\varphi$ and $\varphi'$ (not depicted), are measured relative to the $x$-axes within the respective $x$-$y$ planes. Considered are translation/rotation of the particle along/about the $x$-axis with linear velocity $\vv V$/angular velocity $\vv \Omega$. }
  \label{fig:rot_frames}
\end{figure}

As derived in the main text, the effective nonequilibrium temperatures are determined by the spatial average of the temperature profile $T(\vv r)$ weighted by the viscous dissipation function $\phi^{\theta_\perp, x_\perp}(\vv r)$:
\begin{equation}
  \label{eq:Thbm_perp}
  T_\HBM^{\theta_\perp, x_\perp}
  =
  \frac{
    \int \df \vv r \,
    T(\vv r) \phi^{\theta_\perp, x_\perp}(\vv r)
  }{
    \int \df \vv r \,
    \phi^{\theta_\perp, x_\perp}(\vv r)
  }.
\end{equation}
We calculate the above integrals in terms of spherical coordinates where the polar angle $\theta$ is measured with respect to the particle's symmetry axis. This choice of coordinates is convenient since it reflects the symmetry of the temperature profile $T(r,\theta)$ around the particle as sketched in Fig.~\ref{fig:rot_frames} (solid black reference frame).
Without any loss of generality, we consider translation/rotation of the Janus sphere along/about the $x$-axis, \ie, transverse to the particle's symmetry axis. It is convenient to introduce a second, primed coordinate frame, which is rotated by $\ang{90}$ about the $y$-axis of the first one (see dashed pink frame in Fig.~\ref{fig:rot_frames}).
Within the primed frame, the respective dissipation functions (see main text)
\begin{align}
  \label{eq:app_diss_fct_rotHBM}
  \phi^{\theta_\perp}(r,\theta') 
  &\propto
    r^{-6} \sin^2 \theta',
  \\[0.5em]
  \label{eq:app_diss_fct_trHBM}
  \phi^{x_\perp}(r,\theta')
  &=
    f_s(r) \sin^2\theta' + f_c(r)\cos^2\theta',
\end{align}
then reflect rotational symmetry with respect to the $z'$-axis.
The position vectors, $\vv r$ and $\vv r'$, measured within the respective coordinate systems are related via
\begin{equation}
  \label{eq_rotateFrame}
  \begin{split}
    r' \, \sin\theta' \, \cos\varphi' 
    &= -r \cos\theta,
    \\
    r' \, \sin\theta' \, \sin\varphi' 
    &= r \sin\theta\, \sin\varphi,
    \\
    r' \cos\theta'
    &= r \sin\theta \cos\varphi.
  \end{split}
\end{equation}
One easily verifies that the primed spherical coordinates can be expressed in terms of  $(r,\theta,\varphi)$ as
\begin{align}
  \label{eq:rotateFrame_solution_r}
    r' &= r,
    \\
    \theta'
  \label{eq:rotateFrame_solution_theta}
    &=
    \arccos( \sin\theta \cos\varphi ),
    \\
    \varphi'
    \label{eq:rotateFrame_solution_phi}
    &=
    \arctan(-\tan\theta\, \sin\varphi).
\end{align}
We are only concerned with Eq.~\eqref{eq:rotateFrame_solution_theta} as it enters the arguments of the dissipation functions $\phi^{x_\perp,\theta_\perp}(r,\theta'(\theta,\varphi))$ via
\begin{align}
  \label{eq:APP_sin2perp}
  \cos^2\theta'
  =
  1 - \sin^2\theta'  
  =
  \cos^2\varphi \, \sin^2\theta.
\end{align}
Exploiting the orthogonality relations of the Legendre polynomials, similar and straightforward calculations as performed for motion parallel to the symmetry axis lead to the results for $T_\HBM^{\theta_\perp}$ and $T_\HBM^{x_\perp}$ as given in the main text.

\bibliographystyle{frontiersinHLTH&FPHY} % for Health, Physics and Mathematics articles
\bibliography{references}

\end{document}